\documentclass[12pt]{article}

\usepackage{a41}
\usepackage{floatfig,epsfig}
\usepackage{rotating}

\newcommand{\la}[1]{\label{#1}}
\newcommand{\re}[1]{\ (\ref{#1})}
\newcommand{\nn}{\nonumber}
\newcommand{\ed}{\end{document}}
\newcommand{\be}{\begin{equation}}
\newcommand{\ee}{\end{equation}}
\newcommand{\ba}{\begin{eqnarray}}
\newcommand{\ea}{\end{eqnarray}}
\newcommand{\baz}{\begin{eqnarray*}}
\newcommand{\eaz}{\end{eqnarray*}}
\newcommand{\bb}{\begin{thebibliography}}
\newcommand{\eb}{\end{thebibliography}}
\newcommand{\ct}[1]{${\cite{#1}}$}

\begin{document}
\initfloatingfigs
\sloppy

~
\vspace{30mm}

\begin{center}
{\Large \bf The role of secondary Reggeons in central meson production}

\vspace{10mm}
N.I.Kochelev$^{a,}$\footnote{e-mail address: kochelev@thsun1.jinr.ru},
T.Morii$^{b,}$\footnote{e-mail address: morii@kobe-u.ac.jp},
B.L.Reznik$^{c,}$\footnote{e-mail address: reznik@dvgu.ru},
A.V.Vinnikov$^{a,c,}$\footnote{e-mail address: vinnikov@thsun1.jinr.ru}

\vspace{10mm}
{\it \small $^a$ Bogoliubov Laboratory of Theoretical Physics,\\
     Joint Institute for Nuclear Research, \\
     Dubna, Moscow region, 141980 Russia\\
$^b$ Faculty of Human Development, Division of Sciences for
Natural Environment\\
and Graduate School of Science and Technology, \\
Kobe University,
Nada, Kobe 657-8501, Japan \\
$^c$ Far Eastern State University, Sukhanova 8, GSP, Vladivostok,
     690600 Russia }
\end{center}

\begin{abstract}
We estimate the contribution of $f_2$ trajectory exchange to
the central $\eta$ and $\eta^\prime$ production.
It is shown that secondary Reggeons may give a large
contribution to processes of double diffractive meson production
at high energies.

\vspace{3mm}
PACS number(s): 12.40.Nn, 13.60.Le, 12.39.Mk
\end{abstract}

\newpage
The Regge theory provides a natural and economical description of
hadron-hadron interactions at high energies and small momentum transfers
\ct{col}.
In this approach the interaction between colliding particles
is described by the exchange of effective particles, i.e. Reggeons.
At high energies, the pomeron with vacuum quantum numbers gives
the dominant contribution to the hadron-hadron total cross sections.
The Reggeons with quantum numbers different from the vacuum ones can
also contribute to the total cross sections and their contribution is
vanishing at very high energies.

An interest in double diffractive processes (DDP) is mediated
by the idea that they can be a pure source of
glueballs \ct{rob}, \ct{cl1}.  Intensive studies of these processes
have been recently performed by WA102 collaboration at CERN.
The mechanism of the central meson
production in DDP at high energies is usually related
to the double pomeron exchange (DPE) \ct{cl1}, \ct{cl00}, \ct{cl2}.
This conclusion
is based on the following observations:

i) if the two-pomeron fusion
contributes dominantly to the central meson production,
one can explain rather weak energy dependence   of the
total production cross sections;

ii) $t$- and azimuthal
dependences of differential cross sections for the most
mesons are consistent with the
two vector-fusion mechanism and with additional assumption that
pomeron transforms as  vector current;

iii) quantum numbers of these effective vector
exchanges, e.g.  flavour, P- and C-parities,
 are the same as for the pomeron.

However, these arguments fail if one looks into
the details of the experimental
data. For example, even in the simplest case of the light pseudoscalar
meson production we find:

i) the observed cross section of $ \eta $ production is larger than
that of the $ \eta ^\prime$ production \ct{wa102},
while the  DPE mechanism
predicts
the opposite: $\sigma_{\eta}<<\sigma_{\eta '}$.
This conclusion  comes from the consideration of
the flavour-singlet structure of the two-pomeron fusion which leads to
the enhancement  of flavour-singlet meson production \ct{koch}.

ii) the cross section of ${\pi}^0$ meson production does not show any
angular dependence in the range $ 0^{\circ}< \varphi < 150^{\circ} $
and shows
a peak (presumably a contribution from
the diffractive $\Delta$ and nucleon
resonance production
\ct{wapriv}) at
$ \varphi = 180^{\circ} $, while the mechanism of two-vector
fusion predicts the  behaviour like $ \sin^2 \varphi $
with a maximum at $ 90^{\circ} $.
Hence, the mechanism of ${\pi}^0$ production is not
consistent with the two-vector fusion at all.

The main goal of this letter is to
underline the importance of secondary Reggeon
exchanges for central meson production. As an example, we
estimate the contribution of the $f_2$ exchange with $P=C=+1$ to the
central production of $\eta$ and $\eta '$ mesons for WA102 kinematics.

Let us consider a two-Reggeon fusion  contribution to central $\eta$,
$\eta^\prime$ productions, as is presented in Fig. 1,
where the
pomeron and $f_2$ Reggeon are taken into consideration.

\begin{figure}[htb]
\centering
\epsfig{file=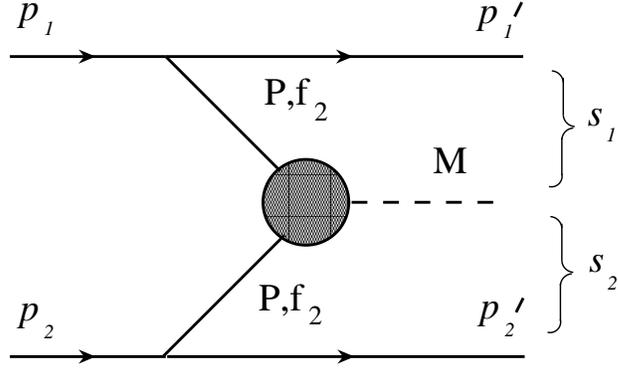,width=0.5\hsize}
\vskip 0.5cm
\caption{ Two Reggeon fusion diagram}
\label{gen}
\end{figure}

The cross section of meson production in the reaction
\be
p(p_1)+p(p_2)\rightarrow p(p_1^\prime)+p(p_2^\prime)+M(p_M)
\nn
\ee
is given by the formula
\be
d\sigma=\frac{dPS^3}{4\sqrt{(p_1.p_2)^2-m_p^4}}
\sum_{spin}|T|^2,
\la{cross}
\ee
where $dPS^3$ is the 3-body phase space volume, $m_p$ is the proton
mass, and $T$ is the matrix element for  DDP reaction.
$\sum_{spin}$ stands for the summation and averaging over
the spins in the final and initial proton states, respectively.

At high energies and small momentum transfers, the four-momenta
of initial and final protons in the center-of-mass system are
given as follows:
\begin{eqnarray}
p_1&\approx&(P+m_p^2/2P,\vec 0,P), {\ }{\ }{\ } {\ }
p_2\approx(P+m_p^2/2P,\vec 0,-P),\nn\\
{p_1}^\prime&\approx&(x_1P+(m_p^2+{\vec{p}_{1T}}^2)/2x_1P
,\vec{p}_{1T},x_1P), \label{kinematic} \\
{p_2}^\prime&\approx&(x_2P+{\vec{p}_{2T}}^2)/2x_2P,\vec{p}_{2T},-x_2P), \nn
\end{eqnarray}
where $P=\sqrt{s}/2$, $s=(p_1+p_2)^2$.
Using the result of Ref. \ct{frix} for the high energy phase
space at small momentum transfers, we obtain
\be
dPS^3=\frac{1}{2^8\pi^4}dt_1dt_2dx_1dx_2d\Phi\delta(s(1-x_1)(1-x_2)-M^2)
\la{ps2}
\ee
for the phase space volume in the DDP reaction,
where $\Phi$ is azimuthal angle between final protons,
$t_{1,2}=(p_{1,2}-p_{1,2}^\prime)^2$ and $M$ is the meson mass.
Kinematic limits for the phase space integration in \re{crossf} are
determined by positive ${\vec{p}_{1,2T}}^2$ in \re{pt} and the
condition $s_{1,2}\geq (M+m_p)^2$, where
\be
s_{1,2}=s(1-x_{1,2}) +
m_p^2 + 2t_{1,2}.  \la{s12}
\ee

Let us consider typical values of $ s_{1,2} $ for the
diffractive process, which appeared to be
very important to understand the reaction mechanism. In the
diffractive region where $t_{1,2}$ are small $ s_{1,2} $ at given
$s$ can be functions of $ x_{1,2}$ only. One can see from (\ref{ps2})
that $ x_{1,2} $ are not independent variables.
At fixed $ x_F=x_2-x_1$
\be
x_1=1-\sqrt{\frac{x_F^2}{4}+\frac{M^2}{s}}-\frac{x_F}{2},~~~
x_2=1-\sqrt{\frac{x_F^2}{4}+\frac{M^2}{s}}+\frac{x_F}{2}.
\la{x12}
\ee
Using (\ref{s12}) and (\ref{x12}) we obtain the dependence of $s_{1,2}$
on $x_F$ (see Fig. 2).
\begin{figure}[htb]
\centering
\begin{minipage}[c]{0.35\hsize}
\epsfig{file=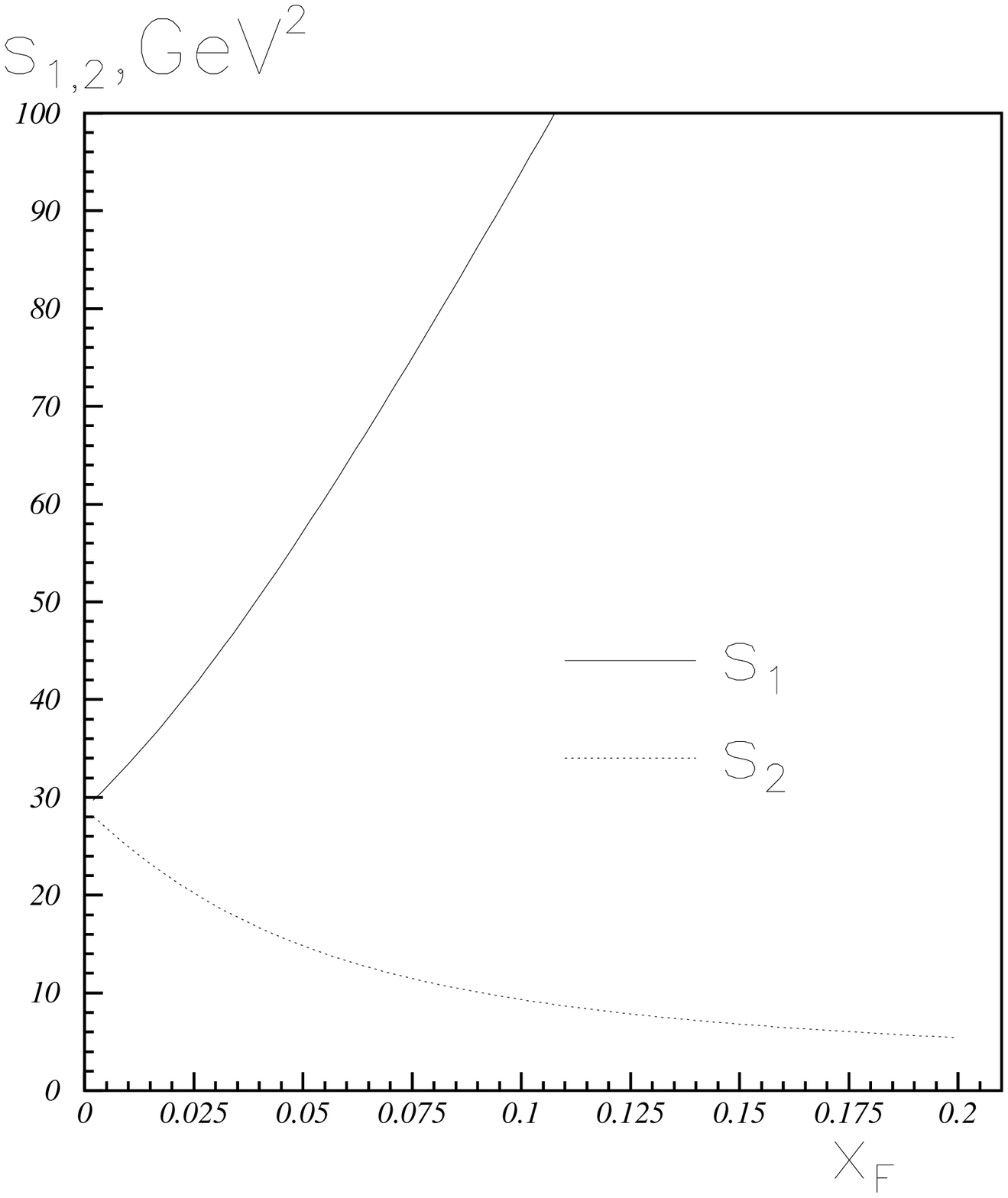,width=\hsize}
\caption{ The dependence of the $s_{1,2} $ on $x_F$ for $\eta ' $ production
at $P$ = 450GeV/c. }
\label{s12x}
\end{minipage}
\hspace*{0.5cm}
\begin{minipage}[c]{0.35\hsize}
\epsfig{file=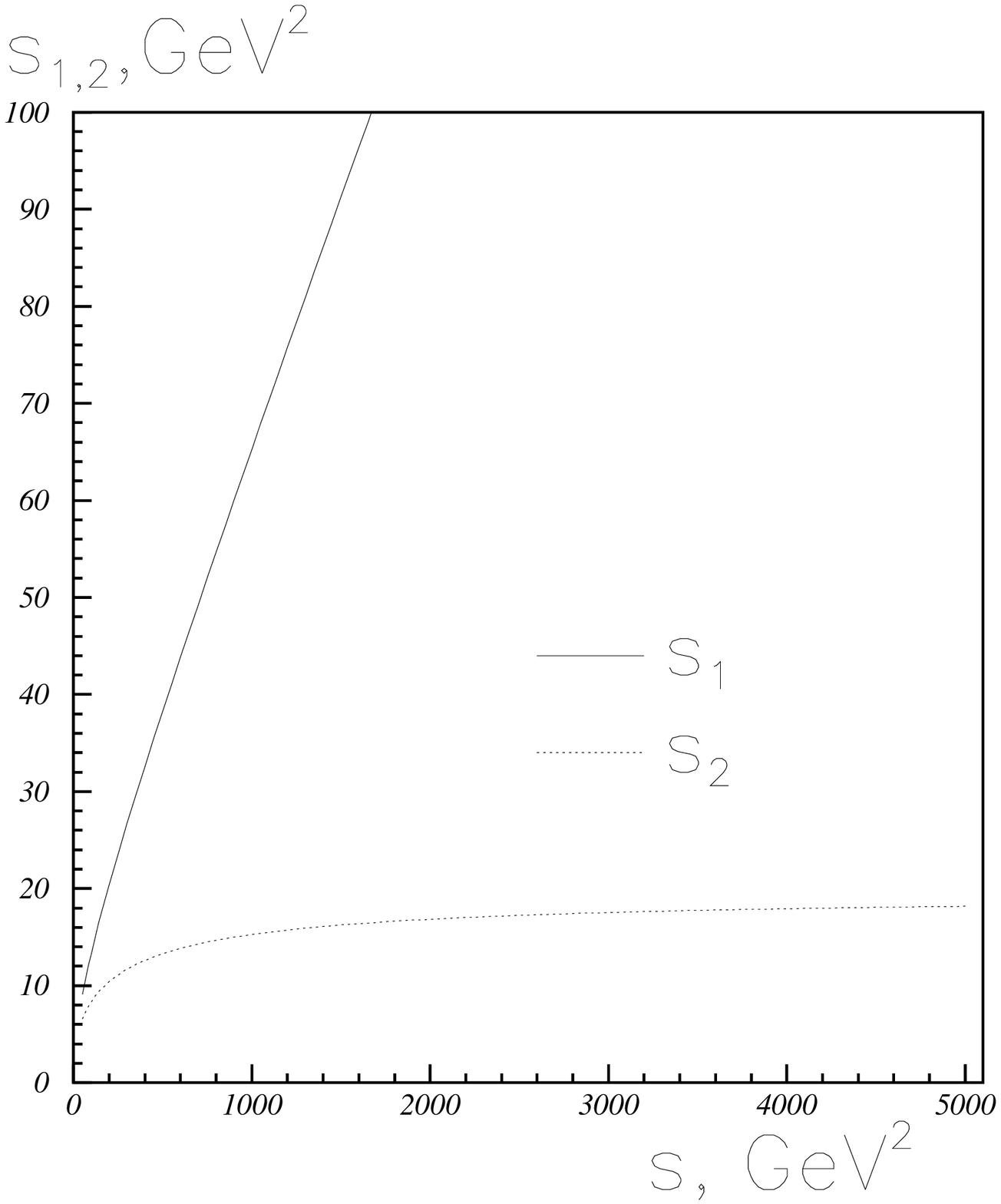,width=\hsize}
\caption{\it The dependence of the $s_{1,2}$ on $s$ for $\eta ' $ production
at $x_F$ = 0.05. }
\label{s12s}
\end{minipage}
\end{figure}
As one can see, despite the large value of the total
invariant mass of the system, the value of $s_2$ is always rather small.
 Therefore,
the large contribution from secondary Regge trajectories can be expected
for any central meson production.
Fixing $x_F$ and varying $s$ (see Fig. 3),
we see that this conclusion remains valid
even at very high energies. This is why even at LHC
energy the contribution from secondary Reggeon exchanges can be not
small, contrary to the expectation \ct{kirk}.

Let us estimate more accurately the contribution of both pomeron and
secondary Reggeons
to the central $\eta, \eta^\prime $ production.
 The leading correction to the two-pomeron fusion  contribution to this
reaction comes from the additional
pomeron-$f_2$ Reggeon fusion, where $f_2$ is the trajectory
with P=C=+1 and positive signature.

The matrix element of DPE reaction, $T$, is given by the formula:
\be
T=i36\beta_P^2\lambda_M A_{PP}^M\epsilon_{\mu\nu\rho\sigma}p_1^\mu p_2^\nu
p_1^{\prime\rho}p_2^{\prime\sigma} F_p(t_1) F_p(t_2) F_{PPM}(t_1,t_2),
\la{tpp}
\ee
where
\be
F_p=\frac{4m_p^2-2.79t}{(4m_p^2-t)(1-t/0.71)^2},
\ee
and
\be
F_{PPM}(t_1,t_2)=\frac{1}{(1-t_1/8\pi^2f_{PS}^2)(1-t_2/8\pi^2f_{PS}^2)}
\ee
are form factors in proton-pomeron and pomeron-pomeron-pseudoscalar
vertices, respectively (see \ct{koch}), and
\be
A_{PP}^M=\beta_P^4 D_{PP}^M\biggl ( \frac{s_1}{s_0} \biggr )^{\alpha_P (t_1)-1}
\biggl ( \frac{s_2}{s_0} \biggr )^{\alpha_P (t_2)-1}
\exp \bigl (-\frac{i\pi}{2}[\alpha_P(t_1)+\alpha_P(t_2)-2] \bigr),
\ee
$s_0=1$ GeV$^2$,
$\beta_P=1.8$ GeV$^{-1}$, $\alpha_P(t)=1+\epsilon+\alpha ' t $ is
the pomeron trajectory with $\epsilon \approx 0.08,~~\alpha ' \approx
0.25$ GeV$^{-2}$ and

\be
\lambda_M = \frac{18}{R_M \alpha_{em}}\sqrt{\frac{2\Gamma_{\gamma \gamma}}
{\pi M^3}}.
\ee
Here $\Gamma_{\eta \rightarrow \gamma \gamma}=0.46\times 10^{-6}$ GeV,
$\Gamma_{\eta \prime \rightarrow \gamma \gamma}=4.28\times 10^{-6}$ GeV
and factors $D_{PP}^M$ and $R^M$ are related to the wave functions of
$\eta$ and $\eta \prime $
\be
\eta=-\sin\Theta\eta_0+\cos\Theta\eta_8, {\ }{\ }
\eta^\prime=\cos\Theta\eta_0+\sin\Theta\eta_8,
\la{wf}
\ee
\be
D_{PP}^{\eta}=-\sin\Theta,{\ }   {\ }   D_{PP}^{\eta^\prime}=\cos\Theta,
\nn
\ee
\be
R_{\eta}=2\sqrt{2}\cos\Theta-\sin\Theta, {\ } {\ }
R_{\eta^\prime}=2\sqrt{2}\cos\Theta+\sin\Theta,
\nn
\ee
where $\Theta=-19.5^\circ$ is the singlet-octet mixing angle.

Using \re{ps2}, \re{x12}, \re{tpp} and the equations
\be
{\vec{p}_{1,2T}}^2=-x_{1,2}t_{1,2}-(1-x_1)^2m_p^2
\la{pt},
\ee
which follow from \re{kinematic},
the cross section is finally written as
\ba
\frac{d\sigma}{dt_1dt_2dx_Fd\Phi}=\frac{3^4\lambda_M^2F_p^2(t_1)F_p^2(t_2)
F_{PPM}^2(t_1,t_2)}{2^9\pi^4\sqrt{x_F^2+4M^2/s}} \nn \\
\times (x_1t_1+(1-x_1)^2m_p^2)
(x_2t_2+(1-x_2)^2m_p^2) |A_{PP}^M|^2 \sin^2\Phi.
\la{crossf}
\ea

The $f_2$ Reggeon  gives an additional contribution to the
total amplitude
\be
A^M=A_{PP}^M+A_{P f_2}^M+A_{f_2 P}^M,
\ee
where
\ba
A_{P f_2}^M=\beta_P^2 \beta_{f_2}^2 D_{P f_2}^M \biggl ( \frac{s_1}{s_0}
\biggr )^{\alpha_P (t_1)-1} \biggl ( \frac{s_2}{s_0} \biggr )^{\alpha_{f_2}
(t_2)-1} \exp \bigl (-\frac{i\pi}{2}[\alpha_P(t_1)+\alpha_{f_2}(t_2)-2]
\bigr), \nn \\
A_{f_2 P}^M=\beta_P^2 \beta_{f_2}^2 D_{P f_2}^M \biggl ( \frac{s_1}{s_0}
\biggr )^{\alpha_{f_2} (t_1)-1} \biggl ( \frac{s_2}{s_0} \biggr )^{\alpha_P
(t_2)-1} \exp \bigl (-\frac{i\pi}{2}[\alpha_{f_2}(t_1)+\alpha_P(t_2)-2]
\bigr).\nn
\ea
The factors $D_{ij}^{M}$ can be obtained from the quark decomposition of
$\eta_1$,
$\eta_8$ and $f_2$ mesons,
\ba
\eta_1=\frac{1}{\sqrt{3}}(u\bar u + d\bar d + s\bar s), \nn \\
\eta_8=\frac{1}{\sqrt{6}}(u\bar u + d\bar d - 2s\bar s), \\
f_2=\frac{1}{\sqrt{2}}(u\bar u + d\bar d ), \nn
\ea
where we assume that $f_2$ is an ideal mixing of $SU(3)$
flavour octet and singlet.

We have
\be
D_{P f_2}^{\eta}= -\sqrt{\frac{2}{3}}\sin\Theta+\frac{1}{\sqrt{3}}
\cos\Theta,~~~D_{P f_2}^{\eta \prime}=\sqrt{\frac{2}{3}}\cos\Theta+
\frac{1}{\sqrt{3}}\sin\Theta .
\ee

The parameters of $f_2$ trajectory have been taken from
Donnachie-Landshoff fit \ct{dl}
\be
\beta_{f_2}=3.6 {\ } \mbox{GeV}^{-1} ,~~~\epsilon_{f_2}=-0.45   ,
\ee
with $\alpha_{f_2}^{\prime}\approx 0.9$ GeV$^{-2}$.

The final result for $\eta$ and $\eta\prime$ production cross sections
for WA102 kinematics ($P=450$ GeV/c, $0<x_F<0.1$) becomes to be equal
to
\be
\sigma(\eta)=450 \mbox{nb},~~~\sigma(\eta\prime)=550 \mbox{nb}.
\ee
which can be compared with the DPE contribution alone \ct{koch}:
\be
\sigma(\eta)=46 \mbox{nb},~~~\sigma(\eta\prime)=370 \mbox{nb},
\ee
and with the experimental data \ct{wa102}
\be
\sigma(\eta)=1295 \pm 120 \mbox{nb},~~~\sigma(\eta\prime)=588\pm 
60 \mbox{nb}.
\ee
We see that $f_2$ Reggeon contribution to the cross section of $\eta$ meson
production  is very large. The admixture
of flavour non-singlet $f_2$ exchange increases the cross section by 
an order of magnitude and also gives the large  enhancement of $\eta  
'$ production.  Taking into account  uncertainties of  our 
model for possible values of quark-pomeron and quark-$f_2$ coupling 
constants as well as representation  of  form factors in  effective 
vertices, we  conclude that the sum of DPE and $f_2$ exchange can 
explain the observed value of total cross section of $\eta$ and 
$\eta\prime$ central production.  On the other hand, the DPE alone 
fails to explain the cross section value.  We should stress 
that $f_2$ exchange does not spoil a good DPE description of the $t$ 
and azimuthal dependence of the differential cross sections of $\eta$ 
and $\eta^\prime$ production \ct{koch}, which  is just fixed by the 
shape of nucleon and pseudoscalar meson form factors and vector-like 
structure of quark-Reggeon vertex.

\begin{figure}[htb]
\centering
\begin{minipage}[c]{0.35\hsize}
\epsfig{file=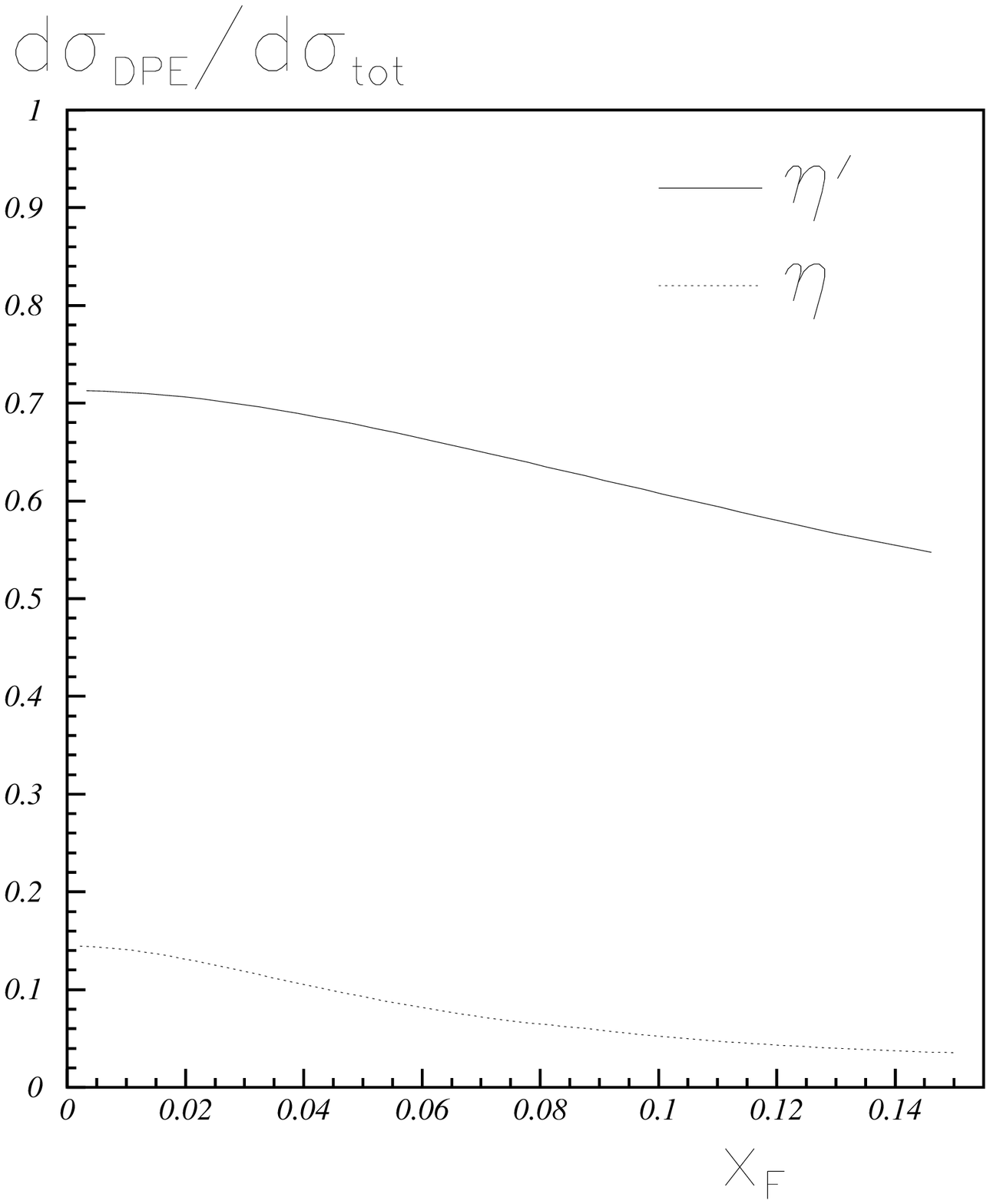,width=\hsize}
\caption{ $d\sigma_{DPE}/d\sigma_{tot}$ as a function of $x_F$.}
\label{relx}
\end{minipage}
\hspace*{0.5cm}
\begin{minipage}[c]{0.35\hsize}
\epsfig{file=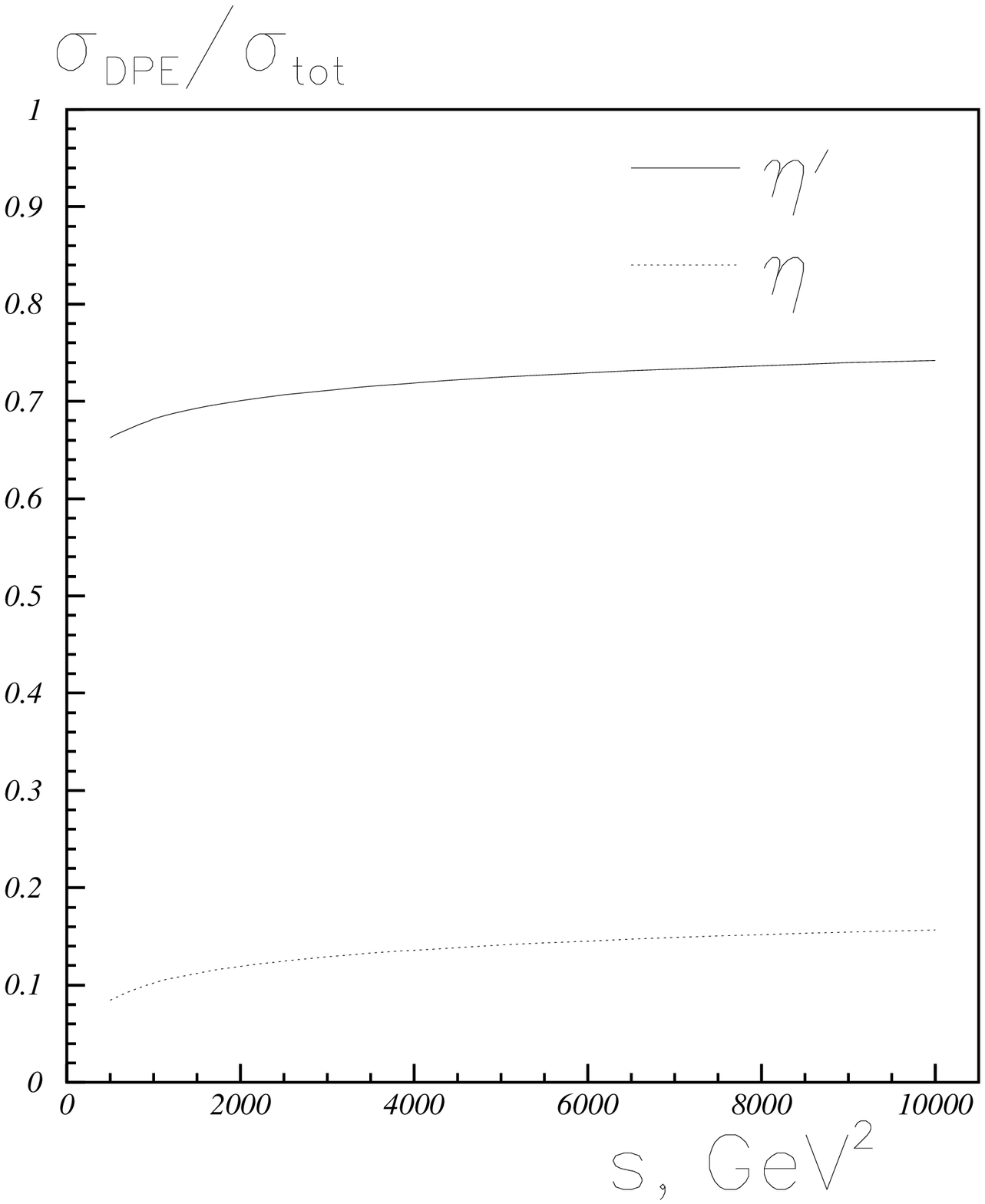,width=\hsize}
\caption{ $\sigma_{DPE}/\sigma_{tot}$ as a function of $s$.}
\label{rels}
\end{minipage}
\end{figure}
We have also analysed the DPE contribution to the cross section
for different values of $x_F$ and $s$ (see Figs. 4 and 5). 
As is seen here, even for
 $\eta\prime$ production, its contribution is not
dominant at any value of $x_F$ and $s$. Increasing energy does not
lead to the prevailing of DPE in $\eta$ and $\eta\prime$ production. 
This is the reason why meson production within DDP does not seem to be 
a pure kinematical region for the dominance of DPE mechanism even at 
 LHC energy \ct{kirk}.

Although our calculation has been performed only for the simplest case 
of pseudoscalar $\eta$ and $\eta^\prime$ meson production, we think 
that any DDP cannot leave the region where
exchanges by secondary Regge trajectories are significant.
This conclusion is based mainly on kinematical arguments; this is why
we expect them to be correct for any central meson production too.
We should also mention that due to large admixture of flavour singlet
component,
$f_2$ exchange can play an important role even in reactions of central
glueball production.

Concluding, we have estimated the contribution of secondary Reggeon 
trajectories into central production of $\eta$ and $\eta '$ mesons. The 
contribution is shown to be very large. Therefore, before to make some 
definite conclusions about properties of the pomeron from the analysis 
of central production data, one should carefully disentangle the 
secondary Reggeon contribution.

We are grateful to A. Dorokhov, S. Gerasimov, V. Romanovsky, N. Russakovich
and A. Titov for useful discussions

\end {document}